\documentclass[conference]{IEEEtran}
\IEEEoverridecommandlockouts

\usepackage{cite}
\usepackage[OT1]{fontenc} 
\usepackage{amsmath,amssymb,amsfonts}
\usepackage{algorithmic}
\usepackage{graphicx}
\usepackage{textcomp}
\usepackage{xcolor}
\usepackage{comment}
\usepackage[ruled,norelsize]{algorithm2e}

\def\BibTeX{{\rm B\kern-.05em{\sc i\kern-.025em b}\kern-.08em
    T\kern-.1667em\lower.7ex\hbox{E}\kern-.125emX}}
\begin{document}

\title{AI-Augmented Metamorphic Testing for Comprehensive Validation of Autonomous Vehicles}
\author{
    \IEEEauthorblockN{Tony Zhang$^{*}$, Burak Kantarci$^{*}$, Umair Siddique$^{**}$}
    \IEEEauthorblockA{
        $^{*}$School of Electrical Engineering and Computer Science, University of Ottawa, Canada\\
        \texttt{\{yzhan117,burak.kantarci\}@uOttawa.ca}\\[2mm]
        $^{**}$reasonX Labs Inc., Ottawa, Canada\\
        \texttt{umair@reasonx.ai}
    }
}

\maketitle

\begin{abstract}

Self-driving cars have the potential to revolutionize transportation, but ensuring their safety remains a significant challenge. These systems must navigate a variety of unexpected scenarios on the road, and their complexity poses substantial difficulties for thorough testing. Conventional testing methodologies face critical limitations, including the oracle problem—determining whether the system’s behavior is correct—and the inability to exhaustively recreate a range of situations a self-driving car may encounter. While Metamorphic Testing (MT) offers a partial solution to these challenges, its application is often limited by simplistic modifications to test scenarios. In this position paper, we propose enhancing MT by integrating AI-driven image generation tools, such as Stable Diffusion, to improve testing methodologies. These tools can generate nuanced variations of driving scenarios within the operational design domain (ODD)—for example, altering weather conditions, modifying environmental elements, or adjusting lane markings—while preserving the critical features necessary for system evaluation.
This approach enables reproducible testing, efficient reuse of test criteria, and comprehensive evaluation of a self-driving system’s performance across diverse scenarios, thereby addressing key gaps in current testing practices.
\end{abstract}


\section{Introduction}
The development and validation of Autonomous Driving Systems (ADS) \cite{zhao2023autonomous} present significant challenges due to the complexity of real-world environments. Unlike traditional software systems with well-defined inputs and outputs, ADS must operate in dynamic, unpredictable environments that demand sophisticated testing methodologies. Further complicating this challenge, many modern ADS architectures operate as black-box systems \cite{9284628}, making their decision-making processes opaque and their outputs difficult to verify comprehensively.

Metamorphic Testing (MT)\cite{chen2018metamorphic} has emerged as a promising approach for validating ADS, particularly in scenarios lacking a definitive test oracle. MT evaluates system behavior by examining invariant relationships between outputs when inputs undergo controlled transformations. However, conventional MT approaches often rely on elementary transformations and limited metamorphic relationships \cite{segura2016survey}. This simplicity can lead to systems that adapt to specific test patterns rather than developing genuine robustness, potentially resulting in models that perform adequately under test conditions but fail to generalize to real-world scenarios \cite{10.1145/3597503.3639191}.

To address these limitations, we propose integrating advanced generative models, specifically Stable Diffusion \cite{rombach2021highresolution}, to enhance MT capabilities in ADS testing. These generative models excel at producing sophisticated, controlled variations in input data while maintaining environmental coherence \cite{10419041}. By incorporating them into the MT framework, we can generate nuanced scenario modifications—such as variations in lighting conditions, weather patterns, and lane configurations—to rigorously evaluate ADS decision-making processes. As illustrated in Figure 1, the Stable Diffusion-XL model is employed to preserve the lane direction in the original photograph while subtly modifying the background on both sides of the lane. This approach aims to generate variations suitable for application as metamorphic testing cases.

Our approach improves test reproducibility by generating consistent yet diverse scenarios and establishing reusable test oracles based on ADS behavioral invariants. Using generative transformations on camera inputs, we evaluate key ADS components like path planning and obstacle detection. By overcoming traditional MT limitations and leveraging modern generative models, our framework provides a more robust method for validating autonomous system safety.

\begin{figure}[h!]
    \centering
    \includegraphics[width=0.48\textwidth]{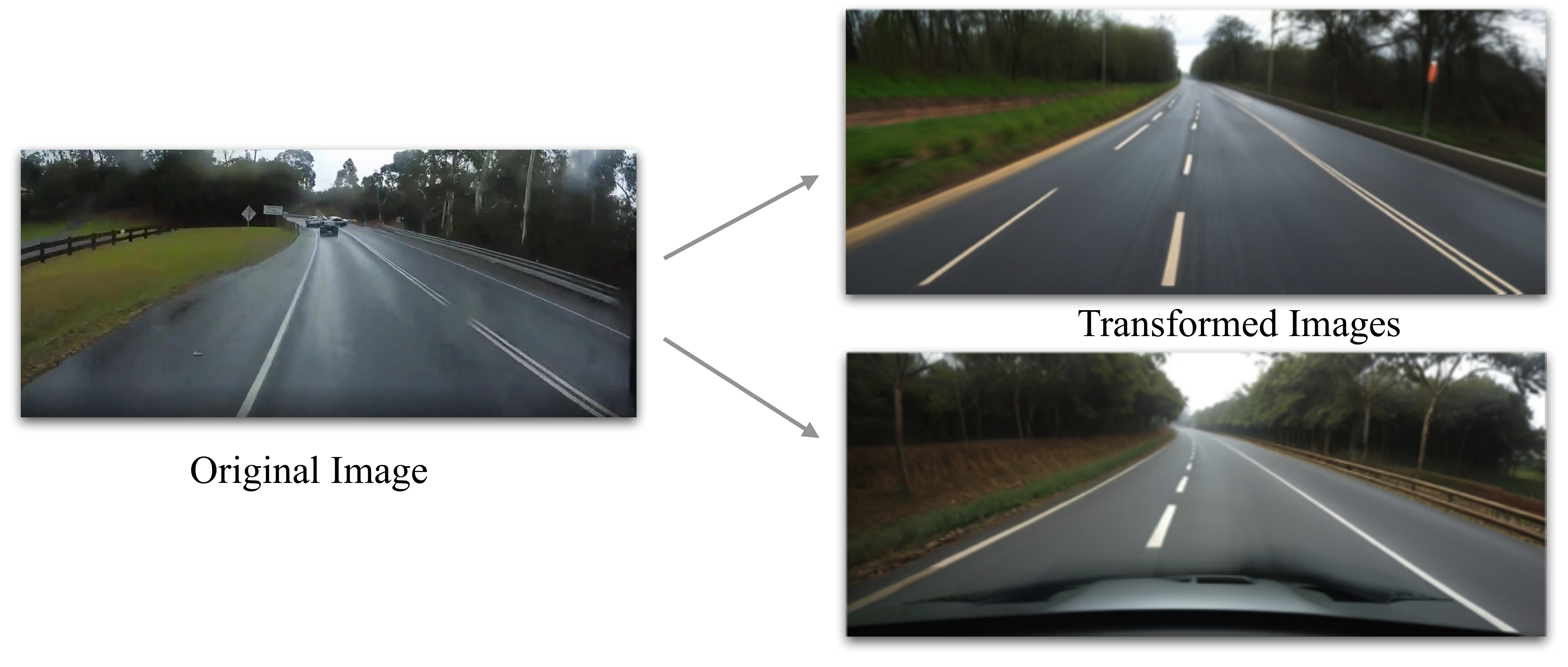} 
    \caption{The case of using Stable Diffusion-XL to slightly change the background of a real image}
\end{figure}

\section{Background}

The validation of Autonomous Driving Systems (ADS) requires careful consideration of both testing methodologies and operational constraints. This section examines three interconnected concepts: Operational Design Domain (ODD) \cite{ISO34503:2023}, Metamorphic Testing (MT), and generative models \cite{oussidi2018deep}.

\subsection{Operational Design Domain}
ODD specifies conditions for ADS functionality, defined as a tuple: $P$ (infrastructure), $E$ (environment), $O$ (constraints), $T$ (temporal), and $C$ (connectivity).

\begin{equation}
    \text{ODD} = (P, E, O, T, C)
\end{equation}

Each component can be further decomposed into specific parameters. For example, the $E$(environment) can be broken down into the following parameters:

\begin{equation}
    \begin{split}
        E = \{&e_w \text{ (weather)}, e_l \text{ (lighting)}, \\
        &e_v \text{ (visibility)}, e_t \text{ (temperature)}\}
    \end{split}
\end{equation}

\subsection{Metamorphic Testing with ODD}
Metamorphic Testing within an ODD framework requires that relations maintain validity within specified operational bounds. For an ADS $S$, input domain $I$, and output domain $O$, we define ODD-constrained metamorphic relations:

\begin{equation}
    \begin{split}
        \text{MR}_{\text{ODD}} \subseteq \{&(x, S(x), x', S(x')) \mid \\
        &x, x' \in I_{\text{ODD}}, \\
        &R(x, S(x), x', S(x')) = \text{true}\}
    \end{split}
\end{equation}

where $I_{\text{ODD}}$ represents inputs valid within the ODD constraints:

\begin{equation}
    I_{\text{ODD}} = \{x \in I \mid \forall c \in \text{ODD}: V(x,c) = \text{true}\}
\end{equation}

Here, $V(x,c)$ verifies compliance with ODD constraint $c$.

\subsection{Generative Models with ODD Integration}
We extend the generative model framework to incorporate ODD constraints. For a generative model $G$ and manually defined transformation specification $\tau$ :

\begin{equation}
    G(x, \tau, \text{ODD}) \rightarrow x' \text{ where } x, x' \in I_{\text{ODD}}
\end{equation}

The transformation specification $\tau$ is now ODD-aware:

\begin{equation}
    \tau_{\text{ODD}} = \{\varepsilon \in E, \gamma \in P, \sigma \in O \times T \times C\}
\end{equation}

This enables the definition of ODD-compliant metamorphic relations:

\begin{equation}
    \begin{split}
        \text{MR}_{G,\text{ODD}} = \{&(x, S(x), G(x,\tau_{\text{ODD}}), \\
        &S(G(x,\tau_{\text{ODD}}))) \mid \\
        &R(x, S(x), G(x,\tau_{\text{ODD}}), \\
        &S(G(x,\tau_{\text{ODD}}))) = \text{true} \wedge \\
        &x, G(x,\tau_{\text{ODD}}) \in I_{\text{ODD}}\}
    \end{split}
\end{equation}

This formulation ensures: 1) Generated scenarios remain within ODD boundaries, 2) transformations preserve ODD-critical properties,  3) test cases maintain operational validity, and
4) validation results are meaningful within the intended operational context.

Integrating ODD with MT and generative models aims to ensure relevant ADS testing within intended conditions while maintaining mathematical validation rigor.

\section{Proposed Approach}

ADSs demand thorough validation within their ODD. We propose a novel framework that integrates MT with generative AI to systematically validate ADS perception systems, addressing three key challenges: 1) Oracle problem \cite{barr2014oracle} in ADS testing, 2) environmental complexity and scenario diversity, 3) uncertainty in perception systems.
\subsection{Framework Overview}

\begin{figure}[h!]
    \centering
    \includegraphics[width=0.48\textwidth]{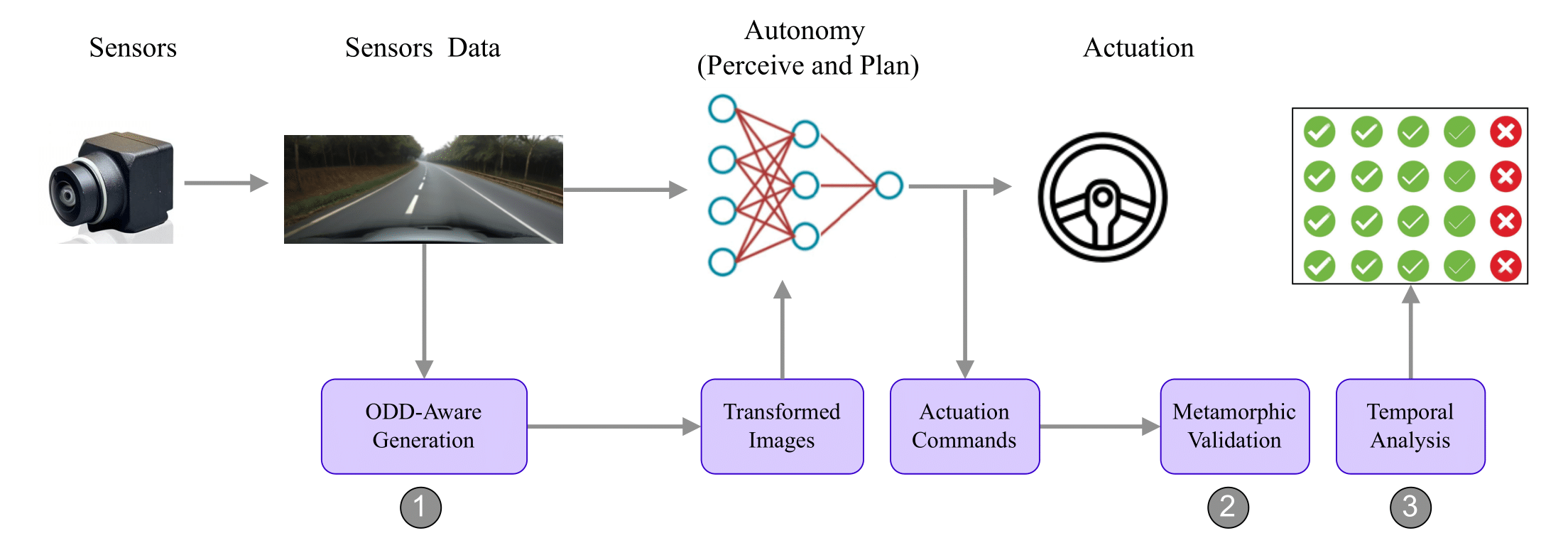} 
    \caption{Architecture of metamorphic testing for ADS and its three key components: (1) ODD-Aware Scenario Generation, (2) Integrated Validation, (3) Time Series Analysis.}
    \label{fig:arch}
\end{figure}

Our framework operates through a systematic workflow (Fig. 2) with two core components:
\begin{itemize}
    \item \textbf{ODD-Aware Generation}: Creates controlled variations in test scenarios while maintaining ODD compliance.
    \item \textbf{Metamorphic Validation}: Evaluates ADS behavior consistency under these variations.
\end{itemize}

Formally, let $S$ represent an ADS under test as follows where $I$ denotes the input space (camera images), $O$ represents the output space (driving decisions), and $h, w$ and $c$ stand for image height, width, and channels, respectively:

\begin{equation}
    S: I \rightarrow O, \text{ where } I \subset \mathbb{R}^{h \times w \times c}
\end{equation}

The framework components are defined as:

\begin{equation}
    \begin{split}
        G_{\text{ODD}}&: I \times \tau \rightarrow I \\
        V_{\text{MR}}&: (I \times O) \times (I \times O) \rightarrow \{0,1\}
    \end{split}
\end{equation}

where $G_{\text{ODD}}$ generates ODD-compliant transformations and $V_{\text{MR}}$ validates metamorphic relations.

\subsection{ODD-Aware Scenario Generation}

\subsubsection{Transformation Space}
For a source image $x \in I$, we define ODD-compliant transformations with the following components:
     Environmental conditions  such as weather and lighting ($\varepsilon$),
    geometric transformations such as perspective and scale ($\gamma$), and
  semantic modifications such as objects, road features ($\sigma$).

\begin{equation}
    \begin{split}
        \tau_{\text{ODD}} = \{&\varepsilon \in E, \gamma \in P, \sigma \in O\} \\
        \text{subject to: } &\forall c \in \text{ODD}: V(G(x,\tau),c) = \text{true}
    \end{split}
\end{equation}

\subsubsection{Generation Process}
involves generation of transformed images that adhere to ODD specifications, utilizing a visual generation model to produce metamorphic testing samples. This is detailed in Algorithm 1, and also illustrated as an integral component (1) of the metamorphic testing architecture for ADS in Fig. \ref{fig:arch}.
\begin{algorithm}[h]
\caption{ODD-Aware Scenario Generation}
\SetAlgoLined
\KwIn{Source image $x$, ODD specifications}
\KwOut{Transformed image $x'$}
Define $\tau$ based on ODD constraints\;
Verify transformation validity: $V(x,\tau) = \text{true}$\;
Generate candidate: $x' \leftarrow G_{\text{ODD}}(x,\tau)$\;
\eIf{ValidateODDCompliance($x'$)}{
    \Return{$x'$}\;
}{
    \Return{GenerateNewTransform($x$)}\;
}
\end{algorithm}

\subsection{Metamorphic Relations and Validation}

\subsubsection{Uncertainty-Aware Relations}
We enhance traditional MRs with uncertainty quantification as follows where $u(\cdot)$ denotes uncertainty quantification, $\theta_u$ represents the uncertainty threshold, and $R(\cdot,\cdot)$ stands for the relation validator:

\begin{equation}
    \begin{split}
        \text{MR}_u(x,x') = \{&(S(x), S(x'), u(S(x)), u(S(x'))) \mid \\
        &R(S(x), S(x')) = \text{true} \wedge \\
        &u(S(x')) \leq \theta_u\}
    \end{split}
\end{equation}

\subsubsection{Validation Criteria}
For each MR category validation criteria are formulated as follows with the following three key components: 
    Path extraction ($P(\cdot)$), Object detection ($D(\cdot)$), and tolerance thresholds ($\epsilon_p, \epsilon_d$).

MR1 and MR2 require that the error threshold is not exceeded, as the generator utilizes similar images for testing. In contrast, MR3 employs an inverse validation relationship, as the generator completely reverses the direction of the lane.

\begin{equation}
    \begin{split}
        V_{\text{MR1}}(x,x') &= \|P(S(x)) - P(S(x'))\| \leq \epsilon_p \\
        V_{\text{MR2}}(x,x') &= \|D(S(x)) - D(S(x'))\| \leq \epsilon_d \\
        V_{\text{MR3}}(x,x') &= P(S(x)) \approx -P(S(x'))
    \end{split}
\end{equation}

\subsection{Temporal Validation}

The third key component of the metamorphic testing framework in Fig. \ref{fig:arch} is time series analysis which aims to ensure robust validation across time sequences as formulated below where   $w$, $\epsilon_t$, and $S_t(\cdot)$ denote time window size, temporal threshold and smoothed prediction, respectively.

\begin{equation}
    \begin{split}
        S_t(x) &= \frac{1}{w}\sum_{i=t-w}^t S(x_i) \\
        V_{\text{temporal}}(x,x') &= \|S_t(x) - S_t(x')\| \leq \epsilon_t
    \end{split}
\end{equation}

\subsection{Integrated Validation Framework}
Integrated validation assesses whether outputs satisfy safety expectations by applying metamorphic relationships, as defined mathematically and outlined in Algorithm 2. This framework serves as the second core component of the metamorphic testing architecture for ADS, illustrated in Fig. \ref{fig:arch}.

\begin{algorithm}[h]
\caption{Integrated Validation Framework}
\SetAlgoLined
\KwIn{Image sequence $X$, ADS $S$, ODD specifications}
\KwOut{Validation report $R$}
Initialize empty report $R$\;
\ForEach{$x_t \in X$}{
    Generate ODD-compliant $\tau_t$\;
    $x'_t \leftarrow G_{\text{ODD}}(x_t, \tau_t)$\;
    $s_t \leftarrow S(x_t)$\;
    $s'_t \leftarrow S(x'_t)$\;
    $u_t \leftarrow$ ComputeUncertainty($s_t$)\;
    $u'_t \leftarrow$ ComputeUncertainty($s'_t$)\;
    $v_{\text{mr}} \leftarrow$ ValidateRelations($s_t, s'_t$)\;
    $v_{\text{temp}} \leftarrow$ TemporalValidation($s_{t-w:t}, s'_{t-w:t}$)\;
    UpdateReport($R, v_{\text{mr}}, v_{\text{temp}}, u_t ,u'_t $)\;
}
\Return{$R$}\;
\end{algorithm}

\subsection{Framework Properties}

The presented framework ensures four essential properties:

\begin{itemize}
    \item \textbf{ODD Compliance:} All transformations respect operational bounds
    \item \textbf{Uncertainty Awareness:} Validation considers prediction confidence
    \item \textbf{Temporal Coherence:} Results remain stable across time
    \item \textbf{Comprehensive Coverage:} Multiple MRs ensure thorough testing
\end{itemize}

This systematic framework enables rigorous validation of ADS perception systems while maintaining practical relevance within specified operational bounds.


\section{Discussions and Ongoing Research}

\subsection{Extension to Multiple Sensor Modalities}
The proposed framework naturally extends to other sensor types:

\begin{itemize}
    \item \textbf{LiDAR Integration:} The metamorphic relations can be adapted for point cloud data as follows     where $p, p'$ represent point clouds and $R_{\text{3D}}$ defines spatial relationships:
    \begin{equation}
        \begin{split}
            \text{MR}_{\text{LiDAR}}(p,p') = \{&(S(p), S(p'), u(S(p)), u(S(p'))) \mid \\
            &R_{\text{3D}}(S(p), S(p')) = \text{true}\}
        \end{split}
    \end{equation}

    \item \textbf{Radar Systems:} Similar principles apply to radar data where where $D_{\text{velocity}}$ extracts velocity measurements:
    \begin{equation}
        V_{\text{radar}}(r,r') = \|D_{\text{velocity}}(S(r)) - D_{\text{velocity}}(S(r'))\| \leq \epsilon_v
    \end{equation}
    
\end{itemize}

\subsection{Reusability Across ADS Platforms}
Our framework’s modular architecture aims to promote reusability in ADS implementations, enabling shared libraries of metamorphic relations and transformations, thus reducing validation overhead across multiple levels:

\begin{itemize}
    \item \textbf{Test Scenarios:} Once generated, transformations can be reused across different ADS versions and configurations
    \item \textbf{Validation Logic:} Metamorphic relations can be encapsulated as reusable components within testing frameworks or pipelines.
    \item \textbf{ODD Specifications:} Formal ODD definitions can be shared across multiple validation campaigns
\end{itemize}

\subsection{Scalability Benefits}

The framework demonstrates remarkable scalability across three critical dimensions:\\
\textbf{Horizontal Scalability:} The system scales seamlessly across autonomous systems, including passenger vehicles, commercial trucks, delivery robots, and industrial AGVs, covering autonomy levels from L2 to L4.

\textbf{Vertical Scalability:} The framework efficiently handles increasing complexity on three fronts:
 1) Single to multi-sensor setups, 2) Simple to complex scenarios, 3) Component to system validation.

\textbf{Operational Scalability:} Testing scales across 1) Development to production, 2) Single to distributed testing, 3) Manual to automated validation.

\subsection{Sensor Fusion Architectures}

The framework's inherent scalability makes it particularly effective for complex sensor fusion architectures \cite{elmenreich2002introduction}, supporting both early and late fusion while ensuring consistent validation. Its modular design enables seamless scaling from single-sensor to multi-sensor systems without requiring major modifications.

\section{Concluding Remarks}
This position paper has introduced a method to transform ADS validation by systematically testing behaviors across diverse scenarios, including rare edge cases. Using reusable metamorphic relations and robust metrics, it aims to enhance safety assurance and fosters innovative testing for more resilient machine learning systems in real-world environments. Our immediate step involves first expanding the diversity of scenarios, including dynamic and adversarial edge cases, and further validating the robustness of the proposed method.

\section*{Acknowledgement}
This work is supported in part by MITACS Accelerate Program under project IT40981 and in part by the NSERC CREATE TRAVERSAL program.
\bibliographystyle{ieeetr}

\vspace{12pt}

\end{document}